# Antiferroelectricity with metastable polar state from the Kittel model

Amit Kumar Shah[1], Xin Li[1*], Guodong Ren[2], Yu Yun[1], Rohan Mishra[2,4], Xiaoshan Xu*[1,3]

[1]Department of Physics and Astronomy, University of Nebraska, Lincoln, NE 68588, USA

[2]Institute of Materials Science and Engineering, Washington University in St. Louis, St. Louis, MO 63130, USA

[3]Nebraska Center for Materials and Nanoscience, University of Nebraska, Lincoln, NE 68588, USA

[4]Department of Mechanical Engineering and Materials Science; Washington University in St. Louis, St. Louis, MO 63130, USA

*Corresponding author: Xiaoshan Xu (X.X.), Xin Li (X.L.)

**Abstract:**

    We have revisited the Kittel model that describes antiferroelectricity (AFE) in terms of two sublattices of spontaneous polarization with antiparallel couplings. By constructing the comprehensive phase diagram including the antiferroelectric, ferroelectric, and paraelectric phases in the parameter space, we identified an AFE phase with stable antipolar states and metastable polar state (SAMP) due to the weak coupling between sublattices. We found that the metastability of the polar state in SAMP phase could lead to apparent remanent polarization, depending on the measurement timescale. This explains the observed ferroelectric behavior of orthorhombic hafnia, which is predicted to be an antipolar by density function theories.

**Introduction:**

Antiferroelectric (AFE) materials may acquire large polarization in an electric field, even though its stable state in zero field has no spontaneous polarization[1-4]. This behavior turns out to be advantageous for capacitive energy storage with high speed and high density[5-7]. Antiferroelectrics have been described phenomenologically by the Kittel model[8], which assumes two neighboring sublattices of spontaneous polarization with antiparallel couplings. Kittel model successfully reproduces the most important features of antiferroelectrics, i.e., zero spontaneous polarization and field-induced large polarization due to the alignment of the two sublattices. Depending on the parameters, Kittel model also predicts, in addition to the antiferroelectric phase, the ferroelectric (FE) phase, as well as the paraelectric (PE) phase.

Orthorhombic-structural hafnia oxide ($HfO_2$), as a material that is compatible with CMOS technology and is promising for integrating ferroelectricity into modern electronics[9-12], matches with the Kittle model on microscopic level[4]. The crystal structure of orthorhombic hafnia consists of well-defined polar layers which are weakly coupled to other polar layers because they are separated by the non-polar spacer layers[13,14], as shown in **Fig.1**(a) for the orthorhombic $Pca2_1$ phase. In addition, density functional calculations predict antiparallel couplings between neighboring polar layers[15], corresponding to an antipolar $Pacb$ phase (see **Fig.1**(b)), while experimental spontaneous polarization has been consistently reported [16-19]. Hence, orthorhombic hafnia may be AFE with weak coupling between sublattices resulting in a stable antipolar states and metastable polar states (SAMP phase), a special case of AFE that hasn't been fully explored in terms of the Kittel's mode.

Like the Landau theory for ferroelectrics with polarization as order parameter, the Kittle model was proposed to describe antiferroelectrics with two order parameters $P_1$ and $P_2$ corresponding to the polarization of two sublattices[1-3,20], as shown in **Fig.1**(c). Under a large enough electric field, the polarization in two sublattice can be aligned, corresponding to antiferroelectric-to-ferroelectric transition (see **Fig.1**(d)). Generally, the free energy can be expressed as:

$$g = \frac{\alpha}{2}(P_1^2 + P_2^2) + \frac{\beta}{4}(P_1^4 + P_2^4) + 2\gamma P_1 P_2 - E(P_1 + P_2) \quad (1)$$

in which $\alpha, \beta$ are Landau coefficients, $\gamma$ is the coupling coefficient, $E$ is the electric field. By reorganizing $P_1$ and $P_2$ to[2]

$$P = \frac{P_1 + P_2}{2}, Q = \frac{P_1 - P_2}{2} \quad (2)$$

the free energy can be rewritten as

$$g(P,Q) = (\alpha + \beta)P^2 + (\alpha - \gamma)Q^2 + \frac{\beta}{2}(P^4 + 6P^2Q^2 + Q^4) - 2EP \quad (3)$$

Here, $P$ and $Q$ are order parameters. Paraelectric (PE), ferroelectric (FE), and antiferroelectric (AFE) states correspond to $\{P = Q = 0\}$, $\{P \neq 0, Q = 0\}$, and $\{P=0, Q \neq 0\}$ as global energy minimum in zero field respectively.

In this work, based on the Kittle model, we show that phase diagram can be determined in terms of the parameters $\alpha, \beta$ and $\gamma$ analytically. Metastable polar and antipolar states are also found as local minima in the free-energy landscape. By investigating the change of free energy landscape under electric field, the field-driven antipolar-to-polar transition can be connected to the polarization switching hysteresis loops qualitatively. In particular, the metastability of the polar state in SAMP AFE phase could cause the apparent remanent polarization up to a finite electric field. Depending on the measurement speed, SAMP AFE phase may exhibit apparent FE loops, which could be related to the ferroelectric loops in HfO$_2$ thin films measured with high frequency.

**Results:**

**Zero field phase diagram from Kittle model**

In the framework of Kittel's model, the thermodynamic stability of various phases is determined by the values of the coefficients $\alpha$ and $\gamma$ in the free energy expansion. To find the stable phases, the energy minimum requires

$$\frac{\partial g}{\partial P} = 2(\alpha + \gamma)P + 2\beta(P^3 + 3PQ^2) - 2E = 0 \quad (4a)$$

$$\frac{\partial g}{\partial Q} = 2(\alpha - \gamma)Q + 2\beta(Q^3 + 3P^2Q) = 0 \quad (4b)$$

and the stability of these phases is determined by the second derivative:

$$D = \frac{\partial^2 g}{\partial P^2}\frac{\partial^2 g}{\partial Q^2} - \left(\frac{\partial^2 g}{\partial P \partial Q}\right)^2 > 0 \quad (5)$$

$$\frac{\partial^2 g}{\partial P^2} = 2(\alpha + \gamma) + 6\beta(P^2 + Q^2) > 0 \quad (6)$$

$$\frac{\partial^2 g}{\partial Q^2} = 2(\alpha - \gamma) + 6\beta(P^2 + Q^2) > 0. \quad (7)$$

The completed phase diagram, as shown in **Fig. 2**(a), can be divided into regions defined by critical conditions such as $\alpha + \gamma = 0$ and $\alpha - \gamma = 0$, as well as other stability-related constraints, including $2\gamma + \alpha = 0$ and $2\gamma - \alpha = 0$. This phase diagram could serve as a detailed map of phase stability and transitions, elucidating the interplay between system parameters and the resulting polarization states. **Fig. 2**(b), (c) and (f) give schematic free energy landscapes of typical PE, AFE, and FE phases, in which there are no metastable states corresponding to local energy minimum. **Fig. 2**(d) shows the schematic free energy landscape for AFE phase with stable antipolar states and metastable polar states (SAMP) states. The schematic energy landscape for FE phase with stable polar states and metastable antipolar states (SPMA) is given in **Fig. 2**(e).

For $\alpha + \gamma > 0$ and $\alpha - \gamma > 0$, which implies $\alpha > 0$, $P = 0$ and $Q = 0$ are the solutions of Eq. (4), corresponding to the paraelectric phase shown in **Fig. 2**(b). In addition, $\frac{\partial^2 g}{\partial P^2} = 2(\alpha + \gamma) > 0$ and $D = 4(\alpha + \gamma)(\alpha - \gamma) > 0$ with $g(P,Q) = 0$ confirms the stability of the paraelectric phase, corresponding to a single-well-like free energy landscape.

For $\alpha + \gamma < 0$ and $\alpha - \gamma > 0$, which implies $\gamma < 0$, Eq. (4) leads to:

$$P = \sqrt{-\frac{\alpha+\gamma}{\beta}} \text{ and } Q = 0, \quad (8)$$

which corresponds to the typical FE phase in **Fig. 2**(f). The stability conditions are satisfied with, $\frac{\partial^2 g}{\partial P^2} = -(\alpha + \gamma) > 0$, $D = 4(\alpha + \gamma)(\alpha + 2\gamma) > 0$. The free energy minimum is $g(P,Q) = -\frac{(\alpha+\gamma)^2}{2\beta} < 0$.

For $(\alpha + \gamma) > 0$ and $(\alpha - \gamma) < 0$, which implies $\gamma > 0$, Eq. (4) leads to:

$$P = 0 \text{ and } Q = \sqrt{-\frac{\alpha-\gamma}{\beta}}, \quad (9)$$

which corresponds to the typical AFE phase in **Fig. 2**(c). Stability is ensured by $\frac{\partial^2 g}{\partial P^2} = -4(\alpha - 2\gamma) > 0$ and $D = 20(\alpha - 2\gamma)(\alpha - \gamma) > 0$. The energy minimum is $g(P,Q) = -\frac{(\alpha-\gamma)^2}{2\beta} < 0$.

For $(\alpha + \gamma) < 0$ and $(\alpha - \gamma) < 0$, Eq. (4) yields four possible solutions. (1) The $P = 0$ and $Q = 0$ solution is unstable as both $\frac{\partial^2 g}{\partial P^2} < 0$ and $\frac{\partial^2 g}{\partial Q^2} < 0$. (2) The antipolar solution ($P = 0$ and $Q = \sqrt{-\frac{\alpha-\gamma}{\beta}}$) is (meta)stable if $D = 16(2\gamma - \alpha)(\gamma - \alpha) > 0$, which leads to $\gamma > \frac{\alpha}{2}$, with a minimum energy $g = -\frac{(\alpha-\gamma)^2}{2\beta}$, and becomes the global minimum if $\alpha\gamma < 0$ ( see **Fig. 2**(d)); otherwise, the antipolar state acts as a saddle point if $\gamma < \frac{\alpha}{2}$. (3) The polar solution ($P = \sqrt{-\frac{\alpha+\gamma}{\beta}}$ and $Q = 0$) is (meta)stable if $D = 16(2\gamma + \alpha)(\alpha + \gamma) > 0$, which leads to $\gamma < -\frac{\alpha}{2}$, with a minimum energy of $g = -\frac{(\alpha+\gamma)^2}{2\beta}$ and the becomes the global minimum when $\alpha\gamma > 0$ [**Fig. 2**(e)]. Conversely, the polar phase acts as a saddle point if $(2\gamma + \alpha) > 0$. (4) The last possible solution ($P = \sqrt{\frac{2\gamma-\alpha}{4\beta}}$ and $Q = \sqrt{\frac{-(2\gamma+\alpha)}{4\beta}}$) is valid if $2\gamma - \alpha > 0$ and $(2\gamma + \alpha) < 0$, ensuring $\frac{\partial^2 g}{\partial P^2} > 0$ and $\frac{\partial^2 g}{\partial Q^2} > 0$, with an energy of $g = \frac{6\gamma^2-\alpha^2}{4\beta}$; however, as $D = 8(2\gamma - \alpha)(2\gamma + \alpha) < 0$, this solution represents saddle points. The detailed calculations are given in the supplementary material.

**Critical electric field for the SAMP AFE phase**

Under an applied electric field, the polar state becomes progressively more stable, whereas the antipolar state becomes less stable. Here, we determine the threshold electric field at which the phase transition occurs. The critical field is calculated systematically through the following steps: (1) Solve Eq. (4b) to obtain the relationship $Q(P)$; (2) Substitute the resulting $Q(P)$ into the stability conditions given by Eqs. (5) and (6) and determine the corresponding values of $P$ that satisfy these conditions; and (3) Insert the derived $P$ and $Q$ values into Eq. (4a) to calculate the critical electric field $E$. The obtained critical fields are

$$E_{10} = \frac{2\gamma + 2\alpha}{3}\sqrt{-\frac{\alpha+\gamma}{3\beta}} \quad (10)$$

$$E_{20} = \frac{4\gamma + 2\alpha}{3}\sqrt{\frac{\gamma-\alpha}{3\beta}} \quad (11)$$

$$E_{40} = \frac{4\gamma - 2\alpha}{3} \sqrt{\frac{2\gamma - \alpha}{3\beta}} \qquad (12)$$

The physical significance of these critical fields depends on the specific parameter conditions, defining clear thresholds for transitions between the antipolar and polar states as shown in the **Fig. S4**. Under the conditions $\gamma > \alpha$ and $\gamma > -\frac{\alpha}{2}$, the zero-field ground state is purely antipolar. Within this parameter range, if $\gamma < \frac{3\alpha}{2}$, the emergence of the polar state and the simultaneous disappearance of the antipolar state occur at the same positive threshold field $E_{20}$. However, if $\gamma > \frac{3\alpha}{2}$, the polar state emerges first at the threshold field $E_{20}$, while the antipolar state disappears subsequently at a higher positive threshold field, $E_{40}$. In the second scenario, where $0 < \gamma < -\frac{\alpha}{2}$, the zero-field state is characterized by a stable antipolar phase coexisting with a metastable polar state. Here, the critical field $E_{20} < 0$ signifies the threshold at which the metastable polar (P+) state disappears under reversed electric fields. The antipolar state disappears at a separate forward-bias threshold field $E_{40} > 0$. For the third condition ($\gamma < 0$ and $\gamma > \frac{\alpha}{2}$), the system at zero field favors a stable polar state alongside a metastable antipolar state. The negative threshold field $E_{10} < 0$ represents the critical field required to remove the metastable polar (P+) state under reverse fields. The antipolar state is eliminated upon applying a positive field exceeding the threshold $E_{40} > 0$. Finally, in the case where $\gamma < \frac{\alpha}{2}$ and $\gamma < -\alpha$, the polar state is stable at zero field. The negative threshold $E_{10} < 0$ corresponds to the critical reverse field at which the polar (P+) phase disappears, reflecting the robustness of the polar ground state under these conditions. This classification clearly defines the conditions for field-induced phase transitions, highlighting when and how antipolar and polar states emerge or vanish under different parameter regimes.

**Fig. 3** presents the variation of the critical electric fields $E_{20}$ and $E_{40}$ as functions of the coupling parameter $\gamma$ for SAMP AFE phase, with fixed coefficients $\alpha = -1.0$ and $\beta = 1.0$. As $\gamma$ increases, the critical field $E_{20}$ decreases, while the critical field $E_{40}$ increases monotonically. Notably, when $\gamma$ approaches $-\frac{\alpha}{2} = 0.5$, $E_{20}$ becomes significantly smaller than $E_{40}$, indicating that the downward polar state (with $P < 0$) becomes unstable first as the electric field is increased. In contrast, as $\gamma$ approaches zero, the two critical fields converge ($E_{20} = E_{40}$), implying a simultaneous disappearance of the antipolar and polar states (with $P < 0$), consistent with a direct and abrupt transition to FE states ($P > 0$). This behavior highlights how the coupling strength $\gamma$ tunes the separation between phase stability thresholds, controlling the nature of the field-driven antipolar-polar transition.

**Field-induced antipolar-to-polar phase transition**

In addition to the critical fields derived above, the field-induced change of free energy landscape provides additional information for the antipolar-to-polar phase transition and vice versa.

First, we consider the typical AFE phase; its schematic position in the phase diagram is given in **Fig. 4**(a). **Fig. 4**(b) and (c) illustrate the electric-field-driven transition between antipolar and polar states for parameters $\alpha = 0, \gamma = 1$, and $\beta = 1$. With this parameter $E_{20} = 0.77$ and $E_{40} = 1.08$. Initially, at lower fields, the antipolar state (orange) is energetically favorable, exhibiting minimal polarization. As the electric field increases, the polar state (blue) emerges first as a local minimum at a critical field $E_{20}$, indicated by the vertical dashed orange line. Upon further increasing the field, the polar state energy continues to decrease and eventually equals that of the antipolar state at an intermediate field (not explicitly labeled). Beyond this point, the polar state becomes energetically favorable (global minimum), and the antipolar state is rendered metastable. Finally, the antipolar state completely disappears at the higher critical field $E_{40}$, marked by the dashed purple line, leaving the polar state dominant. **Fig. 4**(c) clearly shows the corresponding polarization evolution during these transitions.

**Fig. 4**(d) presents contour plots of the free-energy landscape in polarization ($P$) and antipolar order parameter ($Q$) under increasing electric fields, confirming the described phase evolution. Initially, at zero electric field, the landscape exhibits distinct minima at nonzero Q and negligible polarization, indicative of the stable antipolar state. With increasing field, polar minima appear and deepen progressively, leading to coexistence and competition between states. At higher fields ($E > E_{40}$), the polar minima dominate completely, confirming a typical antipolar-to-polar transition characteristic of double hysteresis loops.

In contrast, **Fig. 5** illustrates a special scenario within the antipolar stable region ($\gamma < \frac{3\alpha}{2}$), where the emergence of the polar state and the disappearance of the antipolar state occur simultaneously at the single critical field $E_{20}$, showing no intermediate coexistence or hysteresis. This behavior is characteristic of certain AFE systems exhibiting no hysteresis loops. **Fig. 5**(b) and (c) confirm this simultaneous transition, while **Fig. 5**(d) visually demonstrates the corresponding single-step evolution of energy minimum from antipolar directly to polar stability. As the electric field increases, the small polarization associated with the antipolar state evolves continuously into the larger polarization of the polar state. This smooth and continuous transition, without an intermediate metastable state or energy barrier, results in a single, hysteresis-free polarization switching. Such AFE phases with no hysteresis are particularly promising for energy-storage applications, where minimal energy loss and fast, fatigue-resistant switching are desirable. To our best knowledge, this scenario has been overlooked by earlier work regarding Kittel model [2,3].

Next, we consider the SAMP AFE phase; its schematic position in the phase diagram is given in **Fig. 6**(a). The stability of the polar and antipolar states, along with their polarization behavior under an applied electric field for parameters ($\alpha = -1.0, \gamma = 0.1$, and $\beta = 1$), are illustrated in **Fig. 6**(b) and (c). Initially, the antipolar state represents the global energy minimum, with polar states serving as local minimum (stage 1). As the electric field increases, the energy of the polar ($P < 0$) branch rises, becoming unstable at the critical field $E_{20}$, while the energy of the polar ($P > 0$) and antipolar states decreases. At some intermediate field (not explicitly labeled), the polar ($P > 0$) state achieves equal energy with the antipolar state, beyond which the polar ($P > 0$) state becomes the global minimum, making the antipolar state metastable (stage 2). Further increase of the electric field leads to deeper stabilization of the polar (P>0) state and further destabilization of the antipolar state (stage 3). Finally, when the electric field surpasses the higher critical field $E_{40}$, the antipolar state completely disappears, leaving only the polar (P>0) state as stable (stage 4). **Fig. 6**(c) clearly shows the corresponding polarization evolution, highlighting the critical transitions. Fig. 6(d) provides detailed contours of the free-energy landscape across these four stages, visually demonstrating the evolution of stable and metastable minima with increasing electric field strength.

**Discussion:**

The behavior of the SAMP AFE phase in an electric field may help understand the observed ferroelectric switching in hafnia-based materials on phenomenological level. First, negative domain energy has been shown by the density function theory in the FE hafnia with Pca2$_1$ phase. Phenomenologically, the negative domain wall energy can be written as[15]:

$$E \propto g|\nabla \times P|^2 \qquad (12)$$

in which g<0 is directly related to transverse optic polar phonon, $P$ is the polarization. Given the typical 180-deg domain wall is only separated by the spacer layer within the ac plane, and the polarization P is along $c$ axis, the domain energy can be further simplified to:

$$E \propto g\left|\frac{\partial P_z}{\partial y}\right|^2 = g\left|\frac{P_1 - P_2}{b}\right|^2 = \frac{g}{b^2}(P_1^2 + P_2^2 - 2P_1P_2) \quad (13)$$

in which $P_1, P_2$ are polarization of neighboring polar layers. Obviously, $g<0$ favors antiparallel alignment between $P_1$ and $P_2$, corresponding to a multidomain state of the ferroelectric Pca2$_1$ phase, which is equivalent to the antipolar Pbca phase. In addition, the energy barrier between the polar and antipolar state of hafnia is much larger than their energy difference according to the density function theory[15], suggesting that orthorhombic hafnia can be categorized as an SAMP AFE phase in the phase diagram, where $\frac{\gamma}{\alpha}$ is close to zero.

Interestingly, although pinched hysteresis loops have been often observed in hafnia-based thin films, they are attributed either to defect pinning or electric-field induced transition from the tetragonal phase to the polar Pca2$_1$ phase. If the former mechanism is dominant, the wake-up process eventually converts the pinched loops to more FE-like hysteresis loops. As discussed above, in a high frequency measurement, the polarization switching may result in substantial remanent polarization which resembles that of an FE phase, due to the minor difference of critical field $E_{40}$ and $E_{20}$ under small $\gamma$, as well as the non-equilibrium nature of ferroelectric switching, which prevent the double-hysteresis inherited from the thermodynamic equilibrium.

**Conclusion:**

The classification of ferroelectrics and antiferroelectrics is reexamined by Kittle model, leading to the more detailed phase diagram, in which the antipolar and polar state can coexist as stable and metastable states respectively, corresponding to SAMP AFE phase. The critical electric field for destabilizing polar state and antipolar state as well as the change of free energy landscape under electric field provide the explanation for the partially pinched hysteresis loop. Moreover, diminished difference of two critical fields ($E_{40}$ and $E_{20}$), in near zero $\gamma$ condition, could explain why SAMP AFE phase can exhibit ferroelectric-like switching from high-frequency measurement. These results reconcile the controversy in HfO$_2$ for the FE-like polarization switching loop with the AFE phase (Pbca) as ground state, providing the insights to reveal the intrinsic physic mechanism in single-crystalline HfO$_2$ on phenomenological level.

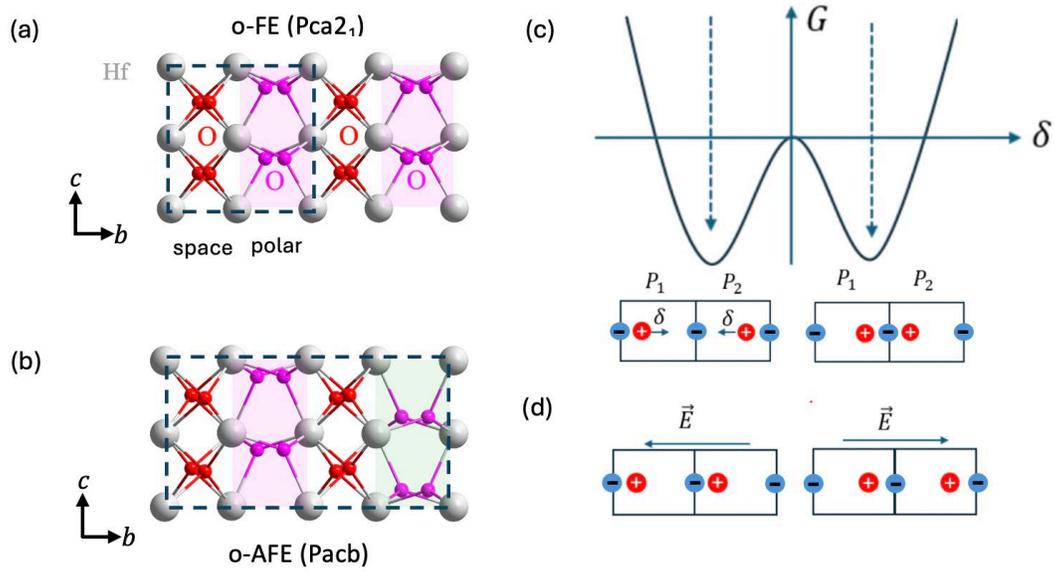

**Fig.1** (a) Atomic structures of ferroelectric Pca2₁ phase and (b) antiferroelectric Pacb phase for HfO$_2$. (c) Schematic free energy profile for antiferroelectrics and related alignment of polarization in two sublattices. (d) The schematic alignment of polarization in antiferroelectrics under electric field.

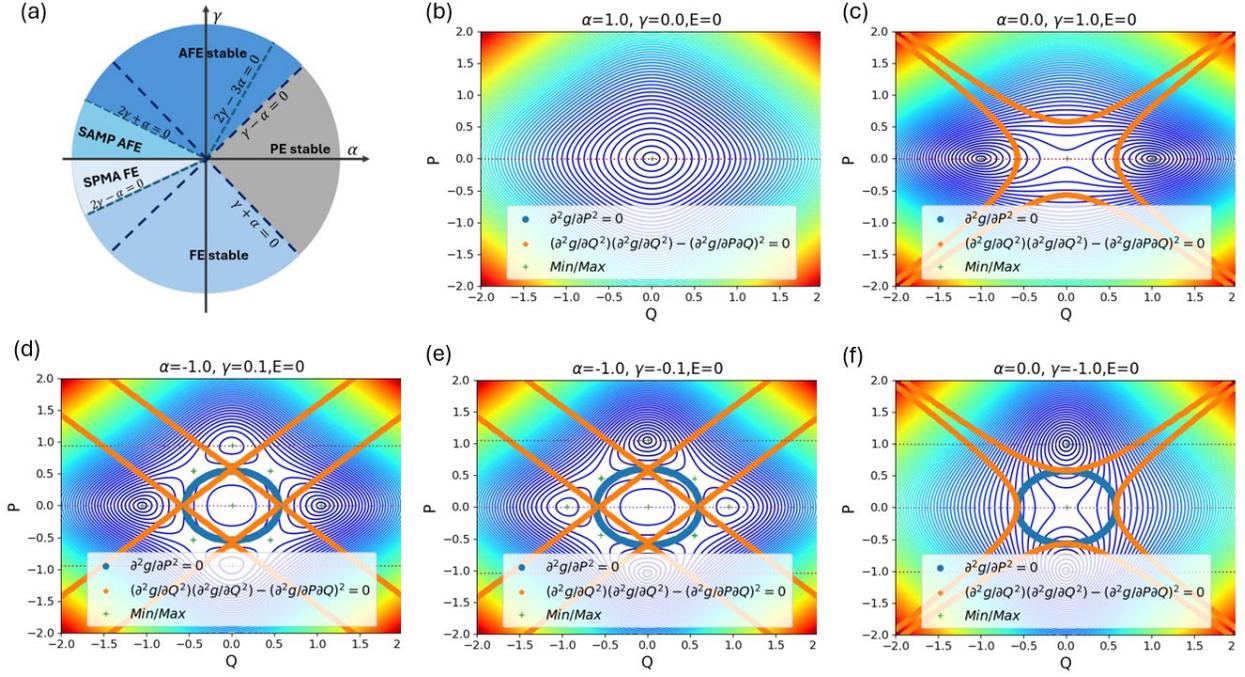

**Fig.2** (a) Phase diagram derived from the Kittle model assuming a constant β. (b) to (f) The energy landscape for typical PE, typical AFE, SAMP AFE, SPMA FE phases, and typical AFE in zero electric field. The orange thick lines indicate $D = \frac{\partial^2 g}{\partial P^2}\frac{\partial^2 g}{\partial Q^2} - \left(\frac{\partial^2 g}{\partial P \partial Q}\right)^2 = 0$. The blue thick lines indicate $\frac{\partial^2 g}{\partial P^2} = 2(\alpha + \gamma) + 6\beta(P^2 + Q^2) = 0$. The symbol "+" indicate positions $\frac{\partial g}{\partial P} = \frac{\partial g}{\partial Q} = 0$.

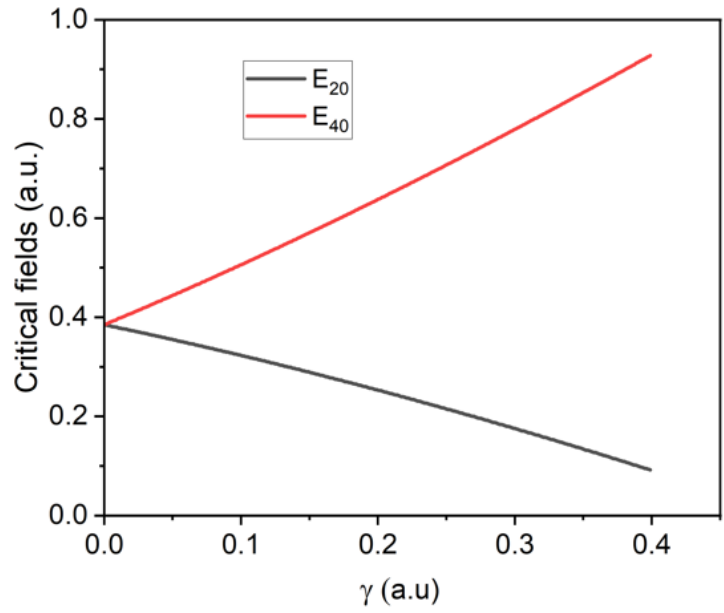

**Fig.3** Critical electric field $E_{20}$ and $E_{40}$ as a function of γ for the SAMP AFE phase.

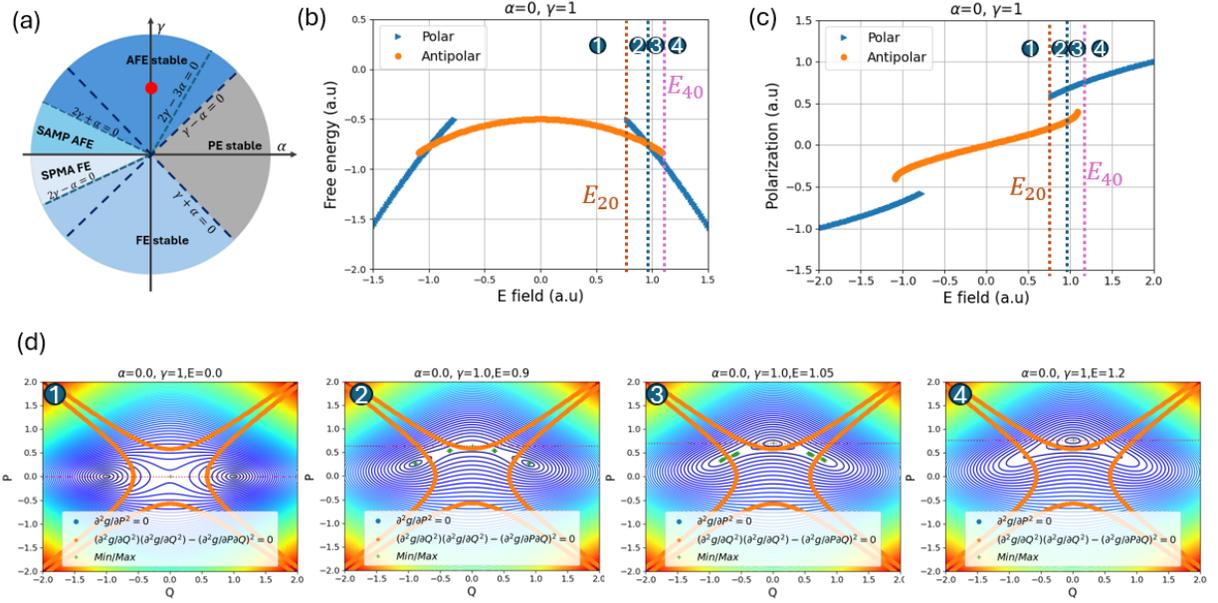

**Fig.4** (a) Schematic position of typical AFE phase on phase diagram. The electric-field-dependent free energy (b) and the related polarization (c) for polar and antipolar states in nonzero *E*. (d) The energy landscape in stages 1 to 4 in (b) and (c).

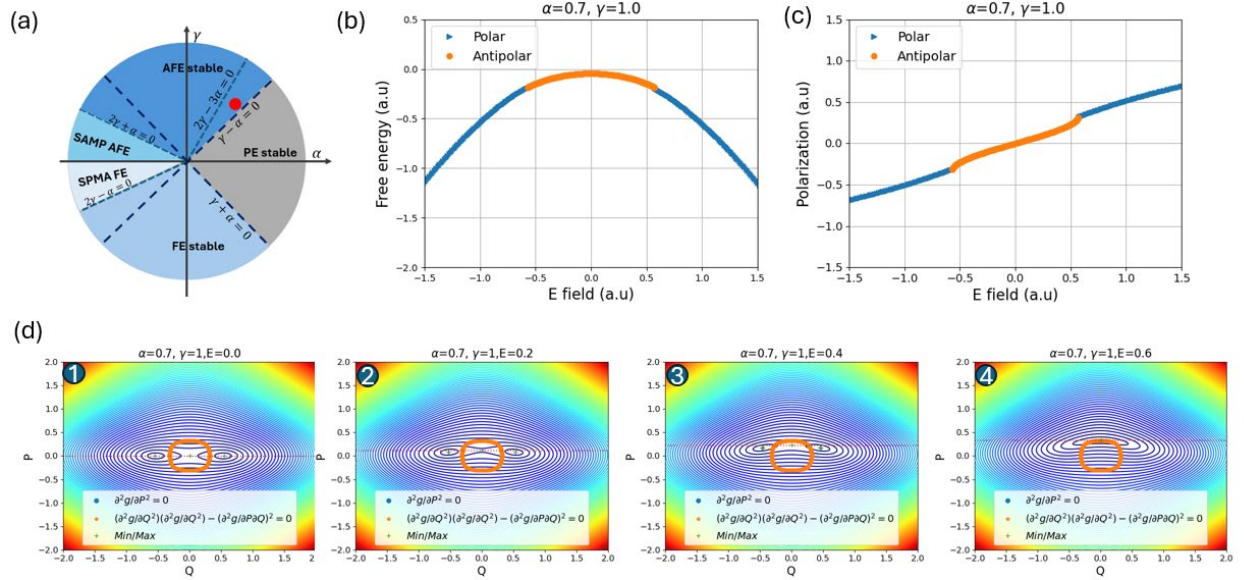

**Fig.5** (a) Schematic position of the AFE phase with no hysteresis. The electric-field-dependent free energy (b) and the related polarization (c) for this state in nonzero *E*. (d) The energy landscape for different fields.

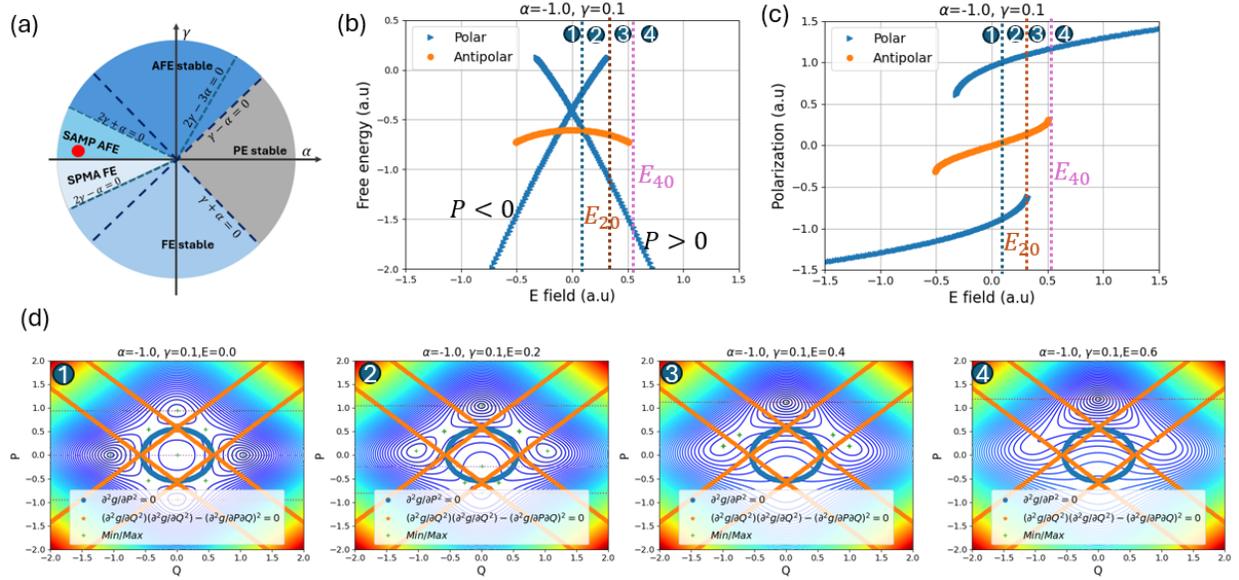

**Fig.6** (a) Schematic position of the SAMP AFE phase on the phase diagram. The electric-field-dependent free energy (b) and the related polarization (c) for the polar and antipolar states in nonzero *E*. (d) The energy landscape in stages 1 to 4 in (b) and (c).